\documentclass[aps,prd,a4paper,nofootinbib,
preprintnumbers,superscriptaddress,showpacs,showkeys,
twocolumn]{revtex4-2}

\usepackage{amsmath}
\usepackage{amssymb}
\usepackage{bm}
\usepackage{graphicx}
\usepackage{xcolor}
\newcommand{\be}{\begin{equation}}
\newcommand{\ee}{\end{equation}}
\newcommand{\ba}{\begin{eqnarray}}
\newcommand{\ea}{\end{eqnarray}}

\begin{document}

%\title{{\textcolor{green}{Exploring the axion potential and domain walls in dense quark matter}}}
%
\title{Exploring the axion potential and axion walls in dense quark matter}

\author{Bonan Zhang}\email{zhangbn@lzu.edu.cn}
\affiliation{School of Nuclear Science and Technology, Lanzhou University, 222 South Tianshui Road, Lanzhou 730000, China}
\affiliation{Frontier Science Center for Rare isotope, Lanzhou University, Lanzhou 730000, China}

\author{David E. A. Castillo}\email{dalvarez@ifj.edu.pl }
\affiliation{Institute of Nuclear Physics, Polish Academy of Sciences, Radzikowskiego 152, 31-342 Cracow, Poland}

\author{Ana G. Grunfeld}\email{grunfeld@tandar.cnea.gov.ar}
\affiliation{CONICET, Godoy Cruz 2290,  C1425FQB Ciudad Aut\'onoma de Buenos Aires, Argentina}
\affiliation{Departamento de F\'\i sica, Comisi\'on Nacional de Energ\'{\i}a At\'omica, Avenida Libertador 8250,
	C1429 BNP, Ciudad Aut\'onoma de Buenos Aires, Argentina}

\author{Marco Ruggieri}\email{marco.ruggieri@dfa.unict.it}
\affiliation{Department of Physics and Astronomy "Ettore Majorana", University of Catania, Via S. Sofia 64, I-95123 Catania, Italy}\affiliation{INFN-Sezione di Catania, Via S. Sofia 64, I-95123 Catania, Italy}

%\vspace{10pt}
%\begin{indented}
%\item[]February 2016
%\end{indented}

\begin{abstract}
We study the potential of the Quantum Chromodynamics 
axion in hot and/or dense
quark matter, within a Nambu-Jona-Lasinio-like model
that includes the coupling of the axion
to quarks. 
Differently from previous studies, we implement local
electrical neutrality and $\beta-$equilibrium,
which are 
relevant for the description of the quark matter in the core  of
compact stellar objects.
Firstly we compute the effects of the
chiral crossover on the axion mass and self-coupling.
We find that the low energy properties of axion are very sensitive
to the phase transition of Quantum Chromodynamics, 
in particular, when the bulk
quark matter is close to criticality.
Then, for the first time in the literature
we compute the axion potential at finite
quark chemical potential and
study the axion domain walls in bulk quark matter.
We find that the energy barrier between
two adjacent vacuum states decrease in the chirally
restored phase: this results in a lower
surface tension of the walls. Finally, we comment on
the possibility of production of walls in
dense quark matter.
\end{abstract}



\maketitle

 %\pagecolor{black}
 
\section{Introduction\label{sec:intro}}

%\textcolor{purple}{Quantum %Chromodynamics (QCD) is the theory 
%that describes strong interactions with  significant features as color confinement, chiral symmetry breaking, and $U(1)_A$ anomaly.} 
%
%
% Quantum Chromodynamics (QCD) is the theory 
% of strong interactions; color confinement,
% chiral symmetry breaking and $U(1)_A$ anomaly 
% are some of its features.
%
%
Quantum Chromodynamics (QCD) is a fundamental quantum field theory that provides a comprehensive framework for describing the strong interaction, which is characterized by a variety of remarkable features including color confinement, chiral symmetry breaking, and the $U(1)_A$ anomaly.
QCD is invariant under gauge transformations
belonging to the $SU(3)$ color group;
however,
gauge invariance does not forbid the term
\begin{equation}
{\cal L}_\theta \propto\theta F\cdot\tilde{F}\label{eq:U1aa}
\end{equation}
in the QCD Lagrangian density. In~\eqref{eq:U1aa},
$F$ and $\tilde{F}$ denote the gluon field strength tensor and its dual respectively, while $\theta$ is a real parameter called the
$\theta-$angle.
A $\theta\neq 0$ would imply an explicit breaking 
of the charge conjugation, $C$, and parity, $P$,
symmetries and QCD would not be invariant under 
$CP$ transformations (thereby inducing an electric dipole moment for the neutron \cite{Crewther:1979pi});
however, there is evidence that $\theta\lesssim 10^{-11}$~\cite{Baker:2006ts, Griffith:2009zz, Parker:2015yka, Graner:2016ses, Yamanaka:2016itb,Guo:2015tla, Bhattacharya:2015esa}.
The fact that $\theta$ is so small 
despite the fact that it is not forbidden by gauge invariance
is called the strong $CP-$problem.
In order to understand this problem
it was suggested  
that a pseudoscalar field, $a$,
exists and couples to the nontrivial
gluon field configurations via the
Lagrangian density
\begin{equation}
{\cal L}_a = \frac{a}{f_a} F\cdot\tilde{F};\label{eq:U1aab_c}
\end{equation}
then, including the $\theta-$term~\eqref{eq:U1aa}
the $CP-$breaking Lagrangian would be
\begin{equation}
{\cal L}_\mathrm{CP} = \theta F\cdot\tilde{F}+\frac{a}{f_a} F\cdot\tilde{F}.\label{eq:U1aab}
\end{equation}
Therefore, violations of $CP$ in strong interactions
would be driven by $\theta + a/f_a$.
The coupling of $a$ to the gluon field gives rise to a potential
for $a$ itself: it was then assumed that this potential develops 
a minimum such that $\langle a/f_a+\theta \rangle= 0$.
Hence, the expectation value of $a$ would cancel the
contributions to observables coming from the
$\theta-$term in Eq.~\eqref{eq:U1aab}. 
This is called the Peccei-Quinn (PQ) mechanism and leads to 
potential solution of the strong $CP-$problem~\cite{Peccei:1977hh, Peccei:1977ur,Weinberg:1977ma, Wilczek:1977pj,GrillidiCortona:2015jxo, Kim:2008hd, Easson:2011zy, Berkowitz:2015aua}.
While the PQ mechanism is quite elegant, it implies the 
existence of a light particle, the QCD-axion (for simplicity
we refer to this particle as the axion in this article),
which represents the quantum fluctuation of the $a$ field
around $\langle a\rangle$.
Axions are dark matter
candidates~\cite{Weinberg:1977ma, Duffy:2009ig, Turner:1990uz, Visinelli:2009zm},
they could arrange in the form of stars~\cite{Tkachev:1991ka,Kolb:1993zz,Chavanis:2011zi,Guzman:2006yc,Barranco:2010ib,Braaten:2015eeu,Davidson:2016uok,Eby:2016cnq,Helfer2017black,Levkov:2016rkk,Eby:2017xrr,Visinelli:2017ooc,Chavanis:2016dab,Cotner:2016aaq,Bai:2016wpg}
and might 
form Bose-Einstein condensates~\cite{Sikivie:2009qn,Chavanis:2017loo}.

Given the wide range in temperature and density 
of the physical systems in which axions might play a role,
it is important to know how the properties
of this particle vary by changing the environment, in particular
temperature and density. 
This is the main scope of the present study, 
in which we compute how the axion properties
are affected by the temperature and the
density of the medium, focusing in particular to 
temperatures and chemical potentials
around the QCD chiral phase transitions.

The use of perturbative methods to study the physical properties of axions around the QCD critical temperature and/or
in dense quark matter is questionable,
hence it is necessary  to resort to QCD-like models and effective field theories to explore physics in the moderate energy scale.
A commomly used effective theory is the Chiral Perturbation Theory ($\chi$PT), which plays an important role in the study of the vacuum structure of QCD as well as the axion properties at low temperatures by means of systematically expanding the 
action in powers of the momenta of
the lighter mesons
\cite{Brower:2003yx,Mao:2009sy,Aoki:2009mx,Bernard:2012ci,Bernard:2012fw,Metlitski:2005di}.
 $\chi$PT
shows great advantages in the low energy regimes, for example, its prediction of the topological susceptibility at zero temperature~\cite{GrillidiCortona:2015jxo}\ is in good agreement with the lattice QCD results~\cite{Borsanyi:2016ksw,Aoki:2017imx,Bonati:2015vqz}. 
However,
at high temperature and/or large density,   $\chi$PT cannot be used 
due to the fact that it lacks information about
the QCD phase transitions. Consequently,
the use of a QCD-like model that is capable to accomodate
axions and the QCD phase transition is very welcome.

In this study,
we use the  Nambu-Jona-Lasinio (NJL) model~\cite{Nambu:1961fr,Nambu:1961tp,Klevansky:1992qe,Hatsuda:1994pi,Buballa:2003qv},
to study the low energy properties of axions.
The model
incorporates the instanton-induced 
interaction that is responsible of the
breaking of the $U(1)_A$ symmetry and 
is capable to describe the spontaneous breaking 
of chiral symmetry as well as the coupling of quarks to the axions. 
In comparison with previous studies, the use of the NJL model
allows us to quantify the effects of the QCD phase transitions
on the low-energy properties of the axion.
We find that the chiral phase transition substantially affects
the axion mass and self-coupling, particularly when
the bulk of dense matter is close to the critical endpoint:
indeed, near the critical endpoint, we find that the axion mass
drops while the self-coupling is enhanced.
Both trends agree with previous model studies
\cite{Lu:2018ukl,Lopes:2022efy,Bandyopadhyay:2019pml,Das:2020pjg,Mohapatra:2022wvj}.

The main novelties of our work can be summarized as follows.
Firstly, we implement $\beta-$equilibrium and
electrical neutrality, keeping in mind
potential applications to compact stellar objects. When compared
to previous works, our approach has the merit to include the
effect of the chiral phase transition on the low-energy properties
of the axions. Secondly, we study the axion walls~\cite{Sikivie:1982qv,Gabadadze:2000vw} and analyze how these
could be produced in the cores of compact stellar objects.
We discuss for the first time how chiral symmetry restoration
in dense quark matter affects the surface tension of the walls.
We then briefly discuss how these walls could form in 
the cores of compact stellar objects.

The plan of the article is as follows. In Section II we present
in some detail the model we use to describe the coupling
of the QCD axion to hot and dense quark matter. In Section III we
present the results on axion mass, self-coupling, potential
and walls at finite temperature and density. Finally, in Section IV
we present our conclusions. Natural units 
$\hbar=1$, $c=1$ and $k_B=1$,
are used throughout the
article.

\section{The model\label{sec:model}}

%\subsection{The Lagrangian density\label{sec:la}}
 %\textcolor{white}{ 
We work in the grand canonical ensemble formalism,
using $T$ and $\mu$ as state variables,
where $\mu$ denotes the quark number 
chemical potential.
We consider quark matter of two light flavors with
Lagrangian density is given by~\cite{Nambu:1961fr,Nambu:1961tp,Klevansky:1992qe,Hatsuda:1994pi,Buballa:2003qv,Ruester:2005jc,Blaschke:2005uj,Lu:2018ukl,Lopes:2022efy,Bandyopadhyay:2019pml,Das:2020pjg,Mohapatra:2022wvj}
\begin{equation}
\mathcal{L} = 
\bar q \left(
i \rlap/\partial
%i\partial\!!\!\!\!\!/ 
+ \hat\mu\gamma_0 - m_0
\right) q +
\bar e \left(
i \rlap/\partial
%i\partial\!!\!\!\!\!/ 
+ \mu_e\gamma_0 
\right) e
+
\mathcal{L}_\mathrm{int}.
\label{eq:Lgene}
\end{equation}
Here $q$ denotes the quark field carrying Dirac,
color and flavor indices, 
while $e$ is the electron field. 
$m_0$ is the current quark mass, that we take to be
equal for $u$ and $d$ quarks for simplicity.
The quark chemical potential 
matrix is
\begin{equation}
\hat\mu = \left(
\begin{array}{cc}
\mu_u & 0 \\
0 & \mu_d
\end{array}
\right)\otimes\bm 1_c
\end{equation}
with $\bm 1_c$ denoting the identity in color space and
\begin{equation}
	\mu_u = \mu-\frac{2}{3}\mu_e,~~~
	\mu_d = \mu + \frac{1}{3}\mu_e;
	\label{eq:Om3}
\end{equation}
$\mu_d=\mu_u + \mu_e$  in agreement with the 
requirement of $\beta-$equilibrium.
Moreover, the interaction term is taken as~\cite{Lu:2018ukl,Lopes:2022efy,Bandyopadhyay:2019pml,Das:2020pjg,Mohapatra:2022wvj}
\begin{eqnarray}
\mathcal{L}_\mathrm{int} &=& 
G_1\left[
(\bar q \tau_a q)(\bar q \tau_a q) +
(\bar q \tau_a i\gamma_5 q)(\bar q  \tau_a 
i\gamma_5 q)
\right] \nonumber \\
&& + 8G_2\left[
e^{i\frac{a}{f_a}}\mathrm{det}(\bar q_R q_L) +
e^{-i\frac{a}{f_a}}\mathrm{det}(\bar q_L q_R) 
\right];
\label{eq:Linte6}
\end{eqnarray}
in particular,
the second line in the above equation corresponds to the
$U(1)_A$-breaking term that is responsible of the coupling
of the QCD-axion to the quarks~\cite{tHooft:1976snw,tHooft:1986ooh}.
In the above equation,
$\tau_a$ are matrices in the flavor
space with
 $a=0,\dots,3$;
$\tau_0$ is the
identity and $\tau_i$ with $i=1,2,3$ 
are Pauli matrices, normalized as 
$\mathrm{tr}(\tau_i \tau_i)=\delta_{ij}/2$.
The coupling constant $G_1$ 
governs the $U(1)_A-$invariant  interaction.
Similarly, $G_2$ regulates the strength
of the $U(1)_A-$breaking term; the determinant
in the latter is understood in the flavor space. 
%} 

%\subsection{Thermodynamic potential\label{sec:ld}}
The thermodynamic potential at one loop 
has been discussed in the literature, see
\cite{Lu:2018ukl} and references therein;
it reads
\begin{equation}
\Omega=\Omega_\mathrm{mf}+ \Omega_{\mathrm{1-loop}} 
+ \Omega_e.
\label{eq:Om1}
\end{equation}
Here we take
\begin{eqnarray}
\Omega_\mathrm{mf}&=&
-G_2(\eta^2-\sigma^2)\cos(a/f_a)
+G_1(\eta^2 + \sigma^2) 
\nonumber\\
&&-2G_2\sigma\eta\sin(a/f_a),
\end{eqnarray}
that represents the mean field contribution to
$\Omega$, with $\sigma=\langle\bar q q\rangle$,
 $\eta=\langle\bar q i\gamma_5q\rangle$.
Moreover,
\begin{equation}
	\Omega_e =
	-
	2 T\frac{4\pi}{8\pi^3} 
	\left(
	\frac{7\pi^4}{180}T^3 + \frac{\pi^2 \mu_e^2 T}{6} + 
	\frac{\mu_e^4}{12 T}
	\right)
	\label{eq:electrons}
\end{equation}
is the contribution of the free, massless electrons.
Finally,
$\Omega_{\mathrm{1-loop}}$ 
corresponds to the quark loop
contribution, given by
\begin{eqnarray}
\Omega_q &=& -4N_c \sum_{f=u,d}\int\frac{d^3p}{(2\pi)^3}
\left[
\frac{E_p}{2} \right.\nonumber\\
&&+\left. 
\frac{1}{2\beta} \log(1+e^{-\beta (E_p-\mu_f)})
(1+e^{-\beta (E_p+\mu_f)})
\right],\nonumber\\
\label{eq:Om2}
\end{eqnarray}
with $\beta=1/T$.
The dispersion laws of quarks are given by
\begin{equation}
E_p = \sqrt{p^2 + \Delta^2},~~~
\Delta^2 = (m_0+\alpha_0)^2 + \beta_0^2,
\label{eq:Om4}
\end{equation}
with
\begin{eqnarray}
&&\alpha_0 = -2\left[G_1 + G_2\cos(a/f_a)\right]\sigma
+2G_2\eta\sin(a/f_a),\nonumber\\
&&\label{eq:Om5_1}\\
&&\beta_0 = -2\left[G_1 - G_2\cos(a/f_a)\right]\eta
+2G_2\sigma\sin(a/f_a).
\nonumber\\
&&\label{eq:Om5_2}
\end{eqnarray}
We notice that the first integral in the
right hand side of Eq.~\eqref{eq:Om2} is
ultraviolet divergent: we regularize this divergence
by cutting the integration at $p=\Lambda$.
The set of parameters we use is \cite{Lu:2018ukl}
$\Lambda=590$ MeV, $G_0\Lambda^2=2.435$,
$G_1=(1-c)G_0$, $G_2=c G_0$, $c=0.2$, $m_0=6$ MeV.

The electron chemical potential
is fixed for each value of the pair $(\mu,T)$
by imposing the electrical neutrality condition
\begin{equation}
\frac{\partial \Omega}{\partial \mu_e}=0.
\label{eq:neut}
\end{equation}
This condition is important for potential applications
to the core of compact stars.
Moreover, the condensates are computed self-consistently
by solving the gap equations
\begin{equation}
\frac{\partial \Omega}{\partial \sigma}=0,~~~
\frac{\partial \Omega}{\partial \eta}=0,
\label{eq:gapeq}
\end{equation}
being sure that the solution $\sigma=\bar\sigma$,
$\eta=\bar\eta$ corresponds to the global minimum
of $\Omega$.

\section{Results\label{sec:resu}}

%\subsection{The neutral ground state\label{sec:ngs}}

\begin{figure}[t!]
	\begin{center}
		\includegraphics[scale=0.23]{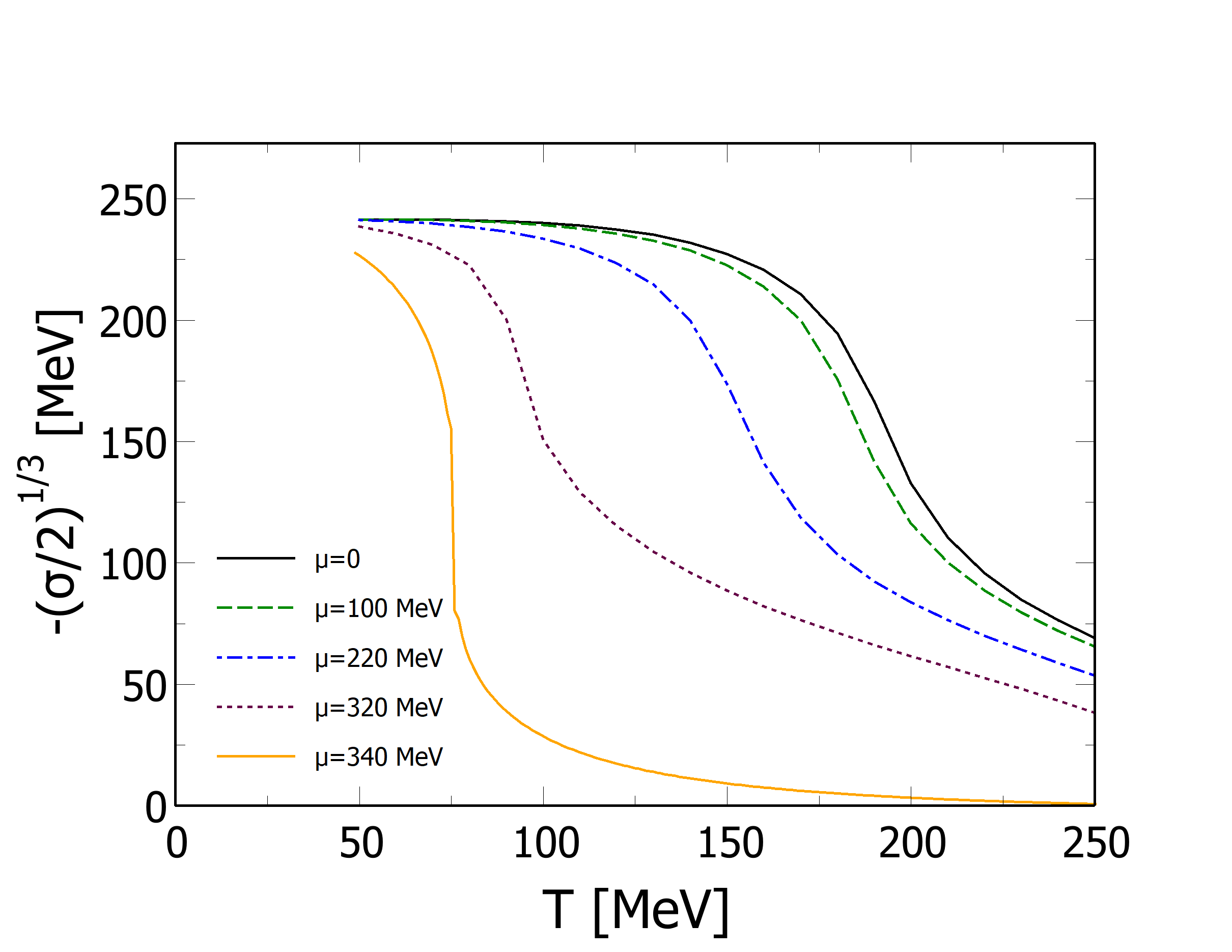}
	\end{center}
	\caption{\label{Fig:cc}
		Chiral condensate,
		-$(\sigma/2)^{1/3}$, versus $T$ for several values
		of $\mu$ in the neutral ground state.}
\end{figure}

In Fig. \ref{Fig:cc} we plot 
$-(\sigma/2)^{1/3}=
-(\langle\bar u u + \bar d d\rangle/2)^{1/3}$
versus $T$ for several values of $\mu$;
the electron chemical potential has been computed
self-consistently by solving simultaneously
the gap equations \eqref{eq:gapeq}
and the neutrality condition \eqref{eq:neut}
for $a=0$. In this case the $\eta-$condensate vanishes.

We notice that for all the values of $\mu$
considered, the chiral condensate drops down
in a narrow range of temperature,
signaling the approximate restoration of
chiral symmetry.
This allows us to define a pseudo-critical
temperature, $T_c$, as the temperature
where $\sigma$ has its largest variation.
$T_c$ drops as the chemical potential increases.
In addition to this,
we notice that the variation of $\sigma$ becomes
sharper with $\mu$: the smooth crossover at $\mu=0$
becomes a sharp transition at
at large $\mu$. 
This implies the existence of a critical endpoint
in the phase diagram: we found it is located at
$(\mu_\mathrm{CP},T_\mathrm{CP})= {(336\textup{MeV},79\textup{MeV})}$.
For completeness, at $T=0$ the critical
chemical potential is $\mu_C=393$ MeV.

%\subsection{The axion sector: $\mu_e=0$\label{sec:axs}}

The axion mass and self-coupling are given by 
\begin{equation}
	m_a^2 =\left.\frac{d^2\Omega}{da^2}\right|_{a=0},
	~~~
	\lambda_a =\left.\frac{d^4\Omega}{da^4}\right|_{a=0},
	\label{eq:ma_1}
\end{equation} 
where the derivatives are total derivatives, namely
they take into account that the two condensates
depend on $a$, and are understood 
at $\sigma=\bar\sigma$,
$\eta=\bar\eta$, where $\bar\sigma$ and $\bar\eta$
are the values of the condensates that minimize
$\Omega$. Since the condensates depend on $a$,
the neutrality condition \eqref{eq:neut} has
to be computed by taking into account this
dependence as well. Thus
\begin{equation}
\frac{d}{da} = \frac{\partial}{\partial a}
+ \frac{\partial\sigma}{\partial a}\frac{\partial}{\partial\sigma}+
\frac{\partial\eta}{\partial a}\frac{\partial}{\partial\eta},
\end{equation}
and so on for the higher derivatives.

\begin{figure}[t!]
	\begin{center}
		\includegraphics[scale=0.3]{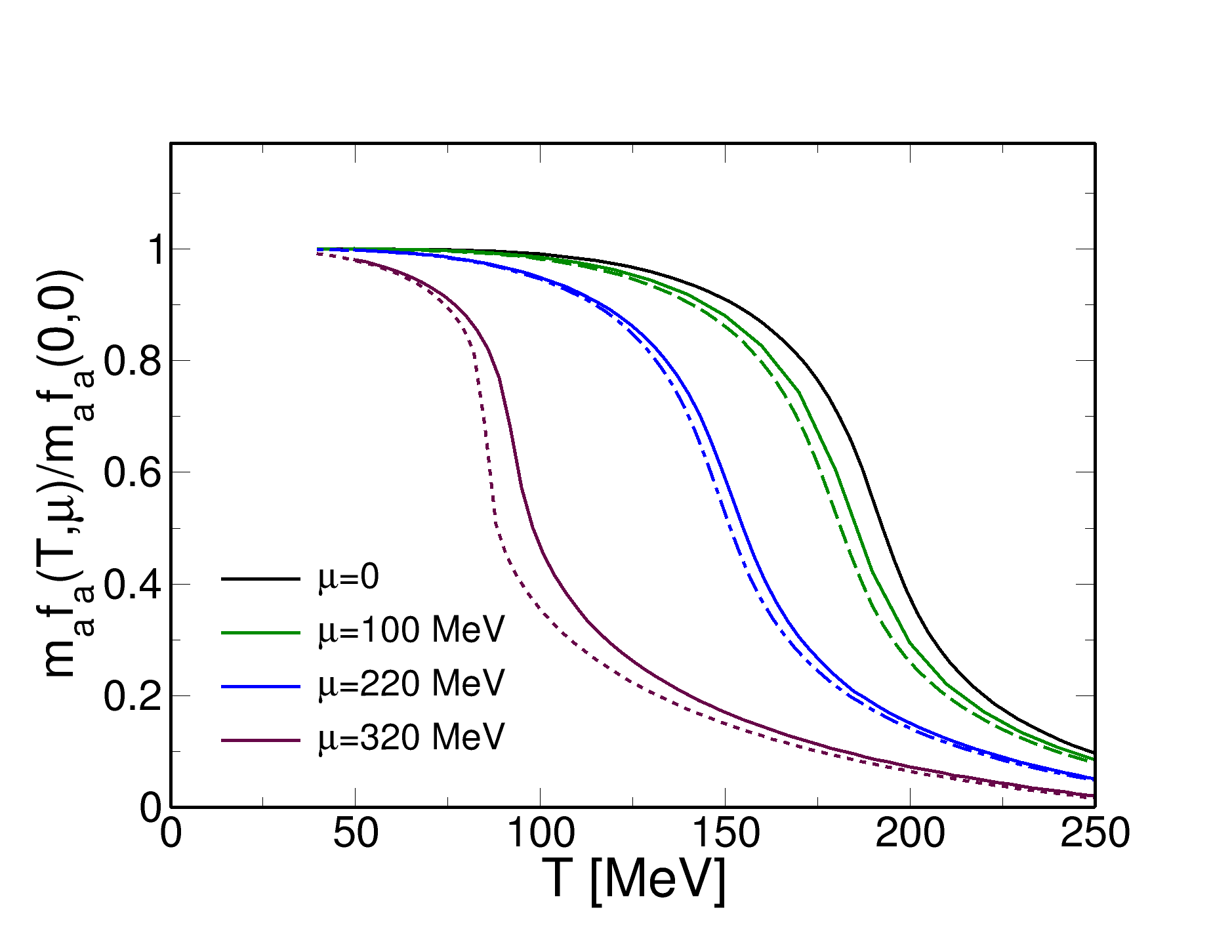}
	\end{center}
	\caption{\label{Fig:axmas2neunorm}
		$m_a f_a$ versus $T$ for several values
		of $\mu$. Solid lines correspond to the
	calculations with electrical neutrality
while dashed lines denote the results for $\mu_e=0$.}
\end{figure}

In Fig.~\ref{Fig:axmas2neunorm} we plot
$m_a f_a$, in units of the same quantity at 
$T=\mu=0$, namely
\begin{equation}
m_a f_a = 6.38\times 10^3~\mathrm{MeV}^2,
\end{equation}
in agreement with previous estimates
\cite{Lu:2018ukl,GrillidiCortona:2015jxo}.
In the figure we plot the results versus temperature, for several values
of $\mu$. The solid lines denote the results
obtained by taking electrical neutrality into
account; for comparison, we show by the dashed lines
the results obtained for $\mu_e=0$.
We note that the decrease
of $m_a$ with $T$
is slightly delayed by $\mu_e\neq 0$; besides this, 
we find no major differences between the cases
with and without the neutrality condition.

From the numerical value of 
$m_a f_a$ in the vacuum we obtain the topological susceptibility,
$\chi=m_a^2 f_a^2$,
which is $\chi\approx (79~\mathrm{MeV})^4$,
again in agreement with previous works~\cite{Lu:2018ukl,GrillidiCortona:2015jxo}.
We notice that in correspondence
of the QCD crossover at finite temperature
the axion mass drops significantly.
Moreover, increasing $\mu$ results in a sharper
drop of the axion mass, similarly to what happens
to the chiral condensate.
We conclude that
the axion mass is very sensitive to the QCD phase transition.

\begin{figure}[t!]
	\begin{center}
		\includegraphics[scale=0.23]{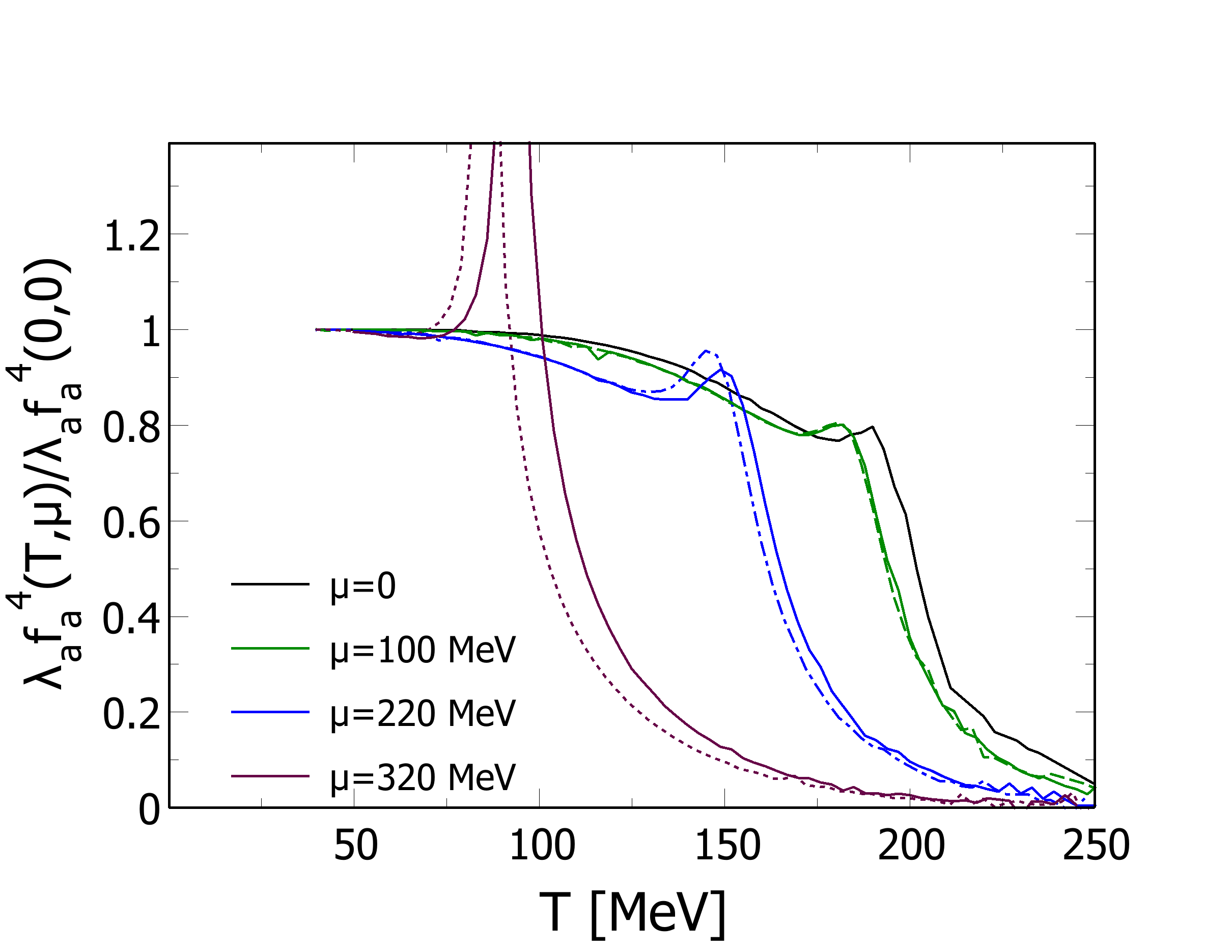}
	\end{center}
	\caption{\label{Fig:qrn2}
		$\lambda_a f_a^4$ versus $T$ for several values
		of $\mu$. Solid lines correspond to the
		calculations with electrical neutrality
		while dashed lines denote the results for $\mu_e=0$.}
\end{figure}

In Fig. \ref{Fig:qrn2} we plot $\lambda_a f_a^4$
versus $T$ for several values of $\mu$.
The solid lines correspond to the results 
obtained by imposing the electrical neutrality
condition while the dashed lines denote  those
with $\mu_e=0$.
At $T=\mu=0$ we find
\begin{equation}
\lambda_a f_a^4= -(55.63~\mathrm{MeV})^4,
\end{equation}
in agreement with previous calculations
within the NJL model \cite{Lu:2018ukl} and with
$\chi$PT \cite{GrillidiCortona:2015jxo}.
The fact that $\lambda_a<0$ means that
the quartic interaction is attractive.
We notice that in correspondence of the chiral
crossover, the quartic coupling experiences
a kink, 
in agreement with~\cite{Mohapatra:2022wvj}; the kink
becomes more pronounced when
the crossover becomes sharper, namely when
the critical endpoint is approached.
Thus, despite the fact that
$\lambda_a$ tends to become smaller with $T$,
the chiral crossover enhances the axion 
self-coupling and this enhancement is very pronounced
in proximity of the critical endpoint.
We also note that imposing electrical neutrality
does not qualitatively change the behavior of $\lambda_a$:
the nonzero $\mu_e$ slightly pushes the
chiral crossover to higher values of $T$; the peaks around the crossover
are still present, and are quite substantial
for large values of the quark chemical potential.

%\subsection{The axion sector: $\mu_e\neq0$\label{sec:axsneu}}

\section{The axion potential and
the domain walls\label{sec:axpot}}

In this section we analyze the full axion
potential \eqref{eq:Om1}, that we later
use to  
analyze the axion domain walls
and in particular to
compute the
surface tension.
The potential $\Omega(\theta)$ in Eq.~\eqref{eq:Om1}
is understood at the global minimum, namely computed
at for the values of $\sigma$ and $\eta$
that minimize $\Omega$ for each value of $\theta$.
In addition to that, since we consider electrically neutral
matter, we fix $\mu_e$ in order to satisfy
the condition  \eqref{eq:neut} for $a \neq 0$.

\begin{figure}[t!]
	\begin{center}
		\includegraphics[scale=0.22]{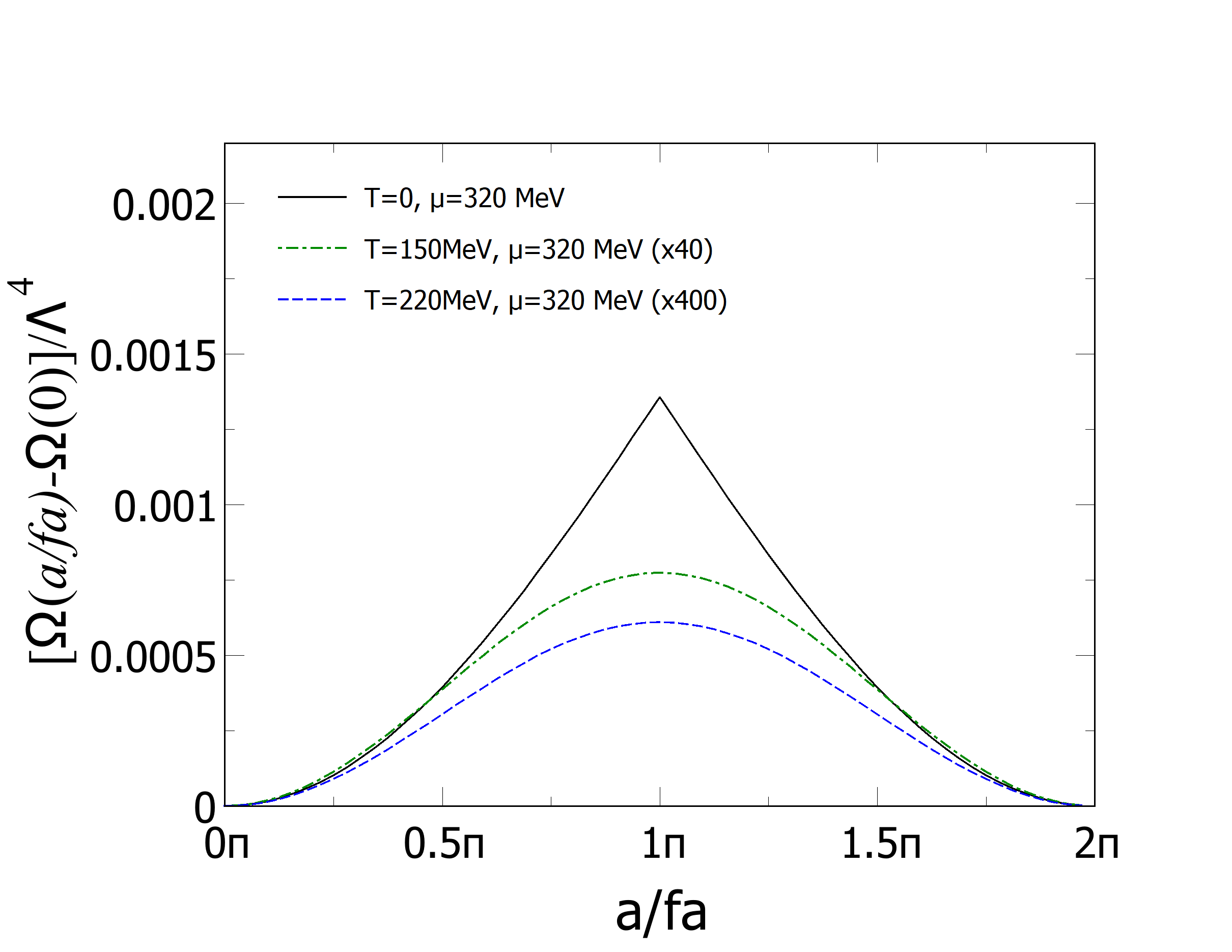}
	\end{center}
	\caption{\label{Fig:axpot320}
		Axion potential at $\mu=320$ MeV,
	computed along the neutrality line.
 The potential is measured in units of the NJL cutoff $\Lambda$.}
\end{figure}

In Fig. \ref{Fig:axpot320} we plot the
axion potential versus $a/f_a$ for several
temperatures and for $\mu=320$ MeV;
this has been computed along the neutrality
line \eqref{eq:neut}.
The value at $a=0$ has been subtracted for later convenience,
see Eq.~\eqref{eq:laax2bef}.
We note that increasing temperature results
the lowering of the potential;
this behavior is in qualitative agreement with
previous results~\cite{Lu:2018ukl,Mohapatra:2022wvj}.
We note that high chemical potential and temperature
the barrier between the two
degenerate vacua $a=0$ and $a/f_a=2\pi$
becomes several orders of magnitude smaller than
that in the phase with chiral symmetry broken.
Consequently, we expect that in the chiral restored
phase the energy stored in solitons connecting the
two vacua will be quite smaller than the one in the vacuum.

As a matter of fact,
the potential shown in Fig. \ref{Fig:axpot320}
gives rise to domain walls that interpolate
between two successive vacua, because the
potential is invariant under
the discrete symmetry transformation 
$\theta\rightarrow \theta + 2\pi n$ with 
$n\in Z$, while this symmetry is broken
spontaneously by choosing one value of $a$,
for example $a=0$.
Derivation of the walls is quite standard
and the details can be found textbooks,
see for example~\cite{Nagashima:2014tva,Shifman:2022shi},
hence here we report the main steps of the
calculations only.

For the domain wall solution
we consider the Lagrangian density
\begin{equation}
	\mathcal{L}=\frac{1}{2}\partial^\mu a 
	\partial_\mu a
	-V(a/f_a),
	\label{eq:laax2bef}
\end{equation}
where we defined
\begin{equation}
	V(x) = \Omega(x)-\Omega(0);
	\label{eq:VdefPPP}
\end{equation}
clearly, $V$ in the above equation depends
on $\mu$ as well, but we suppress this dependence
for the sake of notation.
Incidentally, $V$ is the quantity shown in Fig.~\ref{Fig:axpot320}.
Putting $a=\theta f_a$, the field equation that we get from 
$\mathcal{L}$ is
\begin{equation}
	\partial_\mu\partial^\mu \theta + 
\frac{1}{f_a^2}
	\frac{\partial V(\theta)}{\partial \theta}=0.
	\label{eq:laax4bb}
\end{equation}

The domain wall solution of Eq.~\eqref{eq:laax4bb}
is a solitary wave, 
\begin{equation}
\theta(x,t)=\theta(x-vt),\label{eq:DW_1}
\end{equation}
where $v$ denotes the propagation speed
of the soliton. Putting $\xi=x-vt$ we 
can write Eq. \eqref{eq:laax4bb} as
\begin{equation}
(1-v^2) \theta_{\xi\xi} = \frac{1}{f_a^2}\frac{\partial V(\theta)}{\partial \theta}.
\label{eq:laax4cc}
\end{equation}  
Multiplying both sides of \eqref{eq:laax4cc}
by $\theta_\xi$ and integrating, also noticing that
we impose the boundary conditions
$\theta\rightarrow 0$ and $\theta_\xi\rightarrow 0$
for $\xi\rightarrow\pm\infty$ we have
\begin{equation}
\frac{d\theta}{\sqrt{V(\theta)}}
=
\pm\sqrt{\frac{2}{f_a^2(1-v^2)}}d\xi;
\label{eq:laax4ee}
\end{equation} 
the $\pm$ sign correspond to the kink and
antikink solutions respectively.
The antikink connects $\theta=2\pi$ for $\xi\rightarrow-\infty$
to $\theta=0$ for $\xi\rightarrow+\infty$,
while for the kink the two aforementioned 
limit values of
$\theta$ are inverted.
Equation~\eqref{eq:laax4ee} can be
integrated by 
noticing that, for both the kink and the antikink,
we can
request  that 
in the center of the soliton,
$\xi=0$, 
we have $\theta(\xi)=\pi$. Then
\begin{equation}
	\int_{\pi}^{\theta(\xi)}
	\frac{d\theta}{\sqrt{V(\theta)}}
	=
	\pm \xi \sqrt{\frac{2}{f_a^2(1-v^2)}}.
	\label{eq:laax4eef}
\end{equation} 
The above equation defines implicitly the
soliton $\theta(\xi)$.

In order to simplify the numerical evaluation
of the integral on the left hand side of 
Eq.~\eqref{eq:laax4eef} we represent
the potential by its  
Fourier cosine series,
\begin{equation}
	V(\theta) =
	\frac{c_0}{2} + \sum_{n=1}^N c_n\cos(n\theta),
	\label{eq:serie1}
\end{equation}
with
\begin{equation}
	c_n =
	\frac{2}{\pi}
	\int_0^\pi d\theta~V(\theta)\cos(n\theta).
	\label{eq:serie2}
\end{equation}
For the whole range of $(\mu,T)$ we consider in
this study we find that $N=8$ in 
Eq.~\eqref{eq:serie1} is enough. Moreover,
we find that 
for large $T/\mu$ the approximation $N=1$
works very well, with $c_1=-c_0/2\equiv-V_0$.
For this particular case we have 
\begin{equation}
V(\theta) = V_0(1-\cos\theta)=
m_a^2 f_a^2(1-\cos\theta).
\label{eq:largeTpot}
\end{equation}
For the potential~\eqref{eq:largeTpot} we can perform the integration
in Eq. \eqref{eq:laax4eef} easily to get
\begin{equation}
\theta_\pm(\xi) = 4\arctan\exp\left(
\pm\sqrt{\frac{m_a^2}{ 1-v^2}} \xi
\right),
\label{eq:sol_DW_sine}
\end{equation}
that, besides the rescaling brought by $m_a$,
corresponds to the well-known soliton of the
sine-Gordon equation, propagating along the
$x-$direction with speed $c$.

In the following analysis we consider only
solitons at rest: thus we put $c=0$ in Eq.
\eqref{eq:laax4eef} that implies $\xi=x$ and
\begin{equation}
	\int_{\pi}^{\theta(x)}
	\frac{d\theta}{\sqrt{V(\theta)}}
	=
	\pm x \frac{\sqrt{2}}{f_a }.
	\label{eq:laax4eefbis}
\end{equation} 
For each $x$ 
the above equation allows us to
compute the profile $\theta(x)$; the moving soliton
is obtained from Eq.~\eqref{eq:laax4eefbis} by a
Lorentz boost. Similarly, for the cosine potential~\eqref{eq:largeTpot}
we have
\begin{equation}
	\theta_\pm(x) = 4\arctan\exp\left(
	\pm m_a x
	\right).
	\label{eq:sol_DW_sineBIS}
\end{equation}
The above equation shows that the 
thickness of the wall is 
$\zeta=1/m_a$:
consequently, 
from the results in Fig.~\ref{Fig:axmas2neunorm} we conclude that
chiral symmetry restoration 
(either at high temperature or large baryon density)
results in the
broadening of the axion walls.

\begin{figure}[t!]
	\begin{center}
		\includegraphics[scale=0.22]{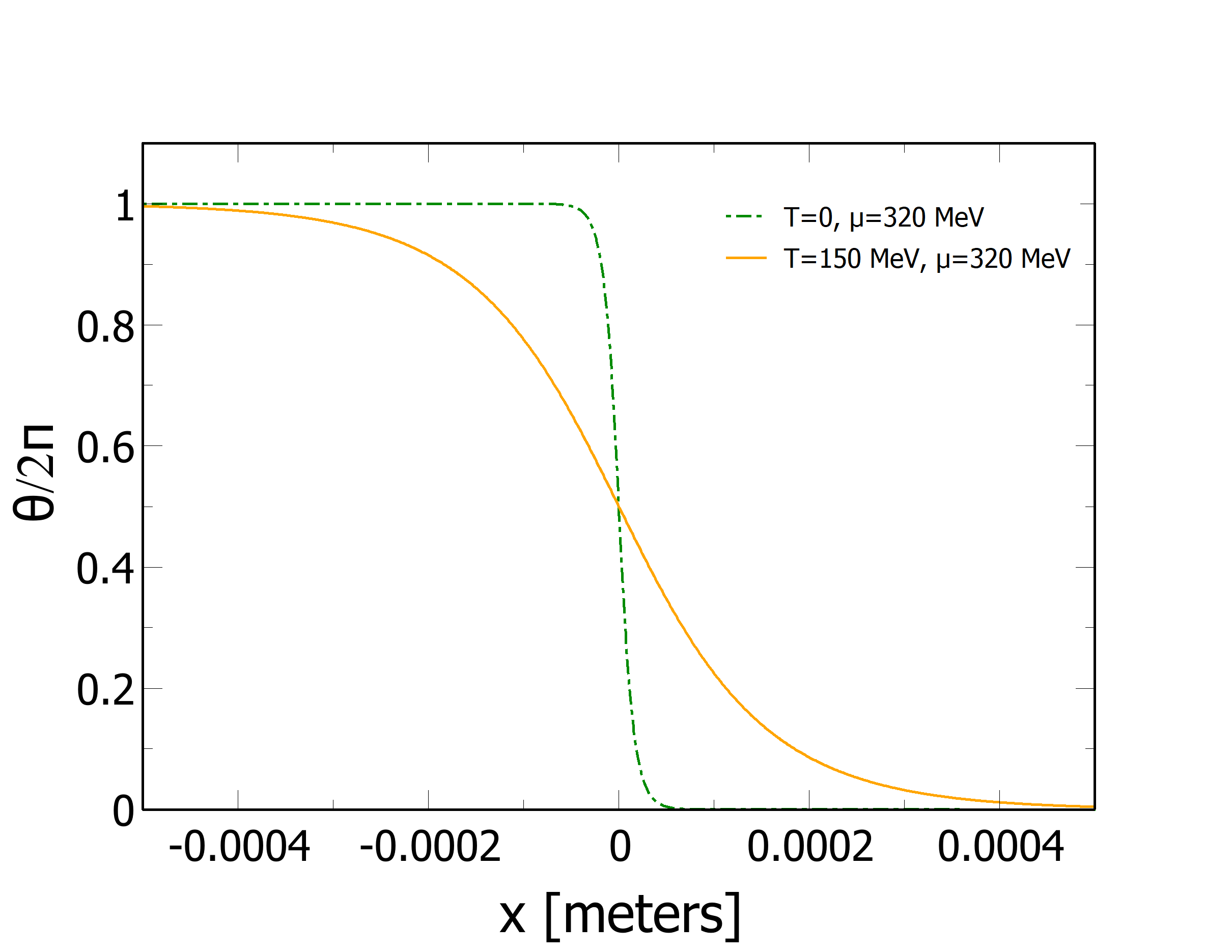}
	\end{center}
	\caption{\label{Fig:profili}
		Axion walls, $\theta=a/f_a$, in the chiral broken phase (green dot-dashed
  line) and chiral symmetric phase (solid orange line).}
\end{figure}

In Fig.~\ref{Fig:profili} we plot the axion wall profiles,
$\theta=a/f_a$,
in the chiral broken phase (green dot-dashed line) and in the chiral
restored phase (orange solid line); 
we used $f_a=10^9$ GeV
which is within the so-called classical
axion window~\cite{Takahashi:2018tdu},
see also~\cite{ParticleDataGroup:2022pth} for more details,
and $m_a$ was computed within the NJL model,
see Fig.~\ref{Fig:axmas2neunorm}:
we found $m_a\approx 6.4$ meV
in the chiral symmetry broken phase and
$m_a\approx 2.4$ meV 
in the chiral symmetry restored phase.
The tiny value of the axion mass in the chiral restored phase explains 
why the spatial extension of the wall in this phase
is of the order of $10^{-4}$ meters.
The qualitative behavior
of the walls is in agreement with the above discussion, namely
restoring chiral symmetry results in the broadening of the walls.
This implies the lowering of the surface tension of the wall,
as we discuss later.

\begin{figure}[t!]
	\begin{center}
		\includegraphics[scale=0.22]{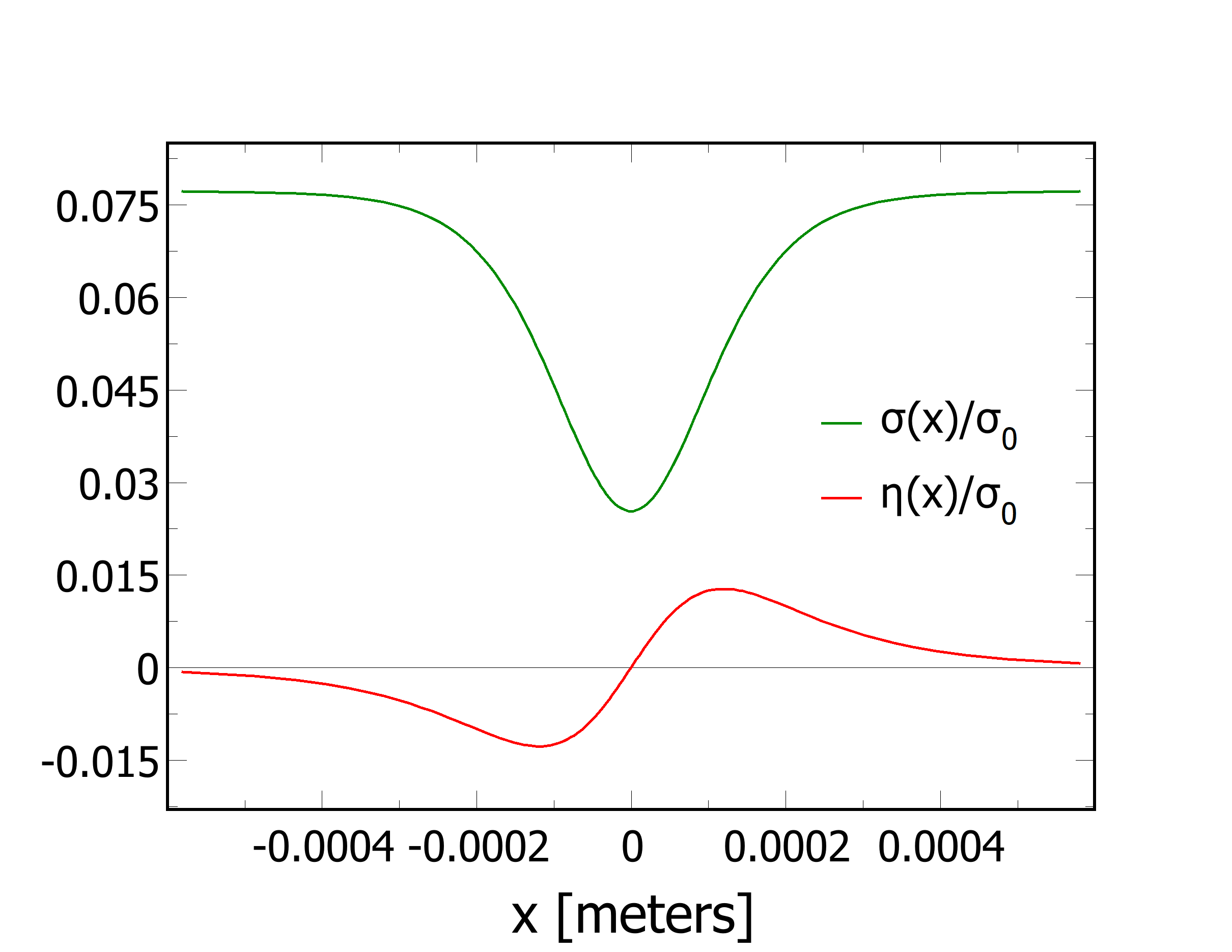}\\
  \includegraphics[scale=0.22]{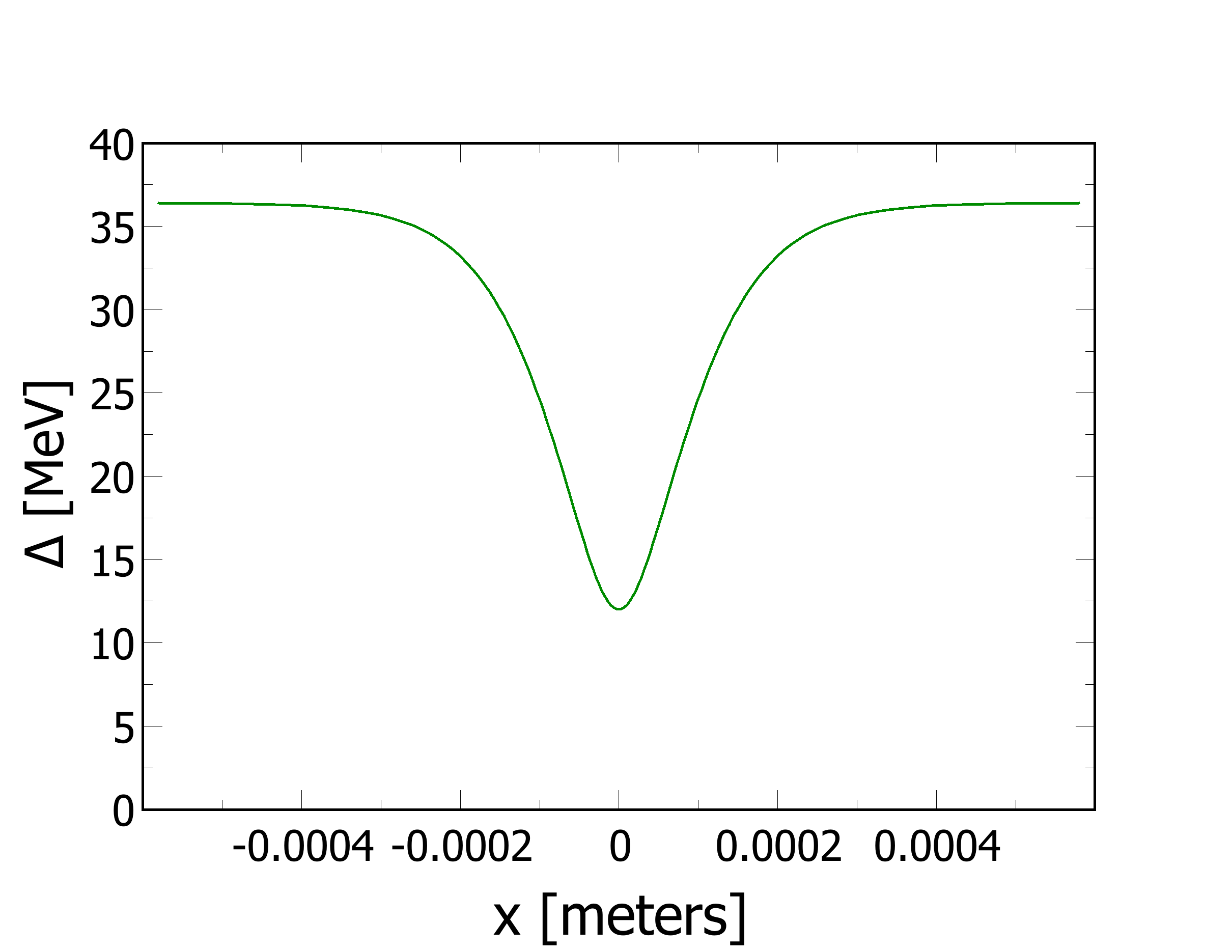}
	\end{center}
	\caption{\label{Fig:structure}
		Condensaltes (upper panel) and
  fermion gap (lower panel) for the wall in cold and dense quark matter.
  We used $f_a=10^9$ GeV, while $m_a=1.48$meV resulting from the NJL model calculation.
  $\sigma_0$ corresponds to the condensate at $T=\mu=0$.}
\end{figure}

It is interesting to analyze the structure of the wall as we 
approach its center.
In the upper panel of
Fig.~\ref{Fig:structure} we plot $\sigma$ and
$\eta$ condensates along an axion wall in the 
cold and dense quark matter phase:
calculations correspond to $\mu=400$ MeV and $T=10$ MeV. 
The condensates are measured in units of 
$\sigma_0=-2 \times (241.5)^3$ MeV$^3$
which corresponds to the condensate in the vacuum.
In the lower panel of the same figure we plot the fermion gap
$\Delta$ defined in Eq.~\eqref{eq:Om4}.
In this phase, the axion potential is well approximated by the
cosine form~\eqref{eq:largeTpot}. In order to compute the condensates
we fixed $f_a=10^{9}$ GeV as before, 
then used the NJL model
to compute $m_a\approx 1.48$ meV.
We note that approaching the core of the wall, 
the $\eta$-condensate forms, signaling the spontaneous
breaking of parity in that region.
We also note that $\Delta$ decreases by a factor of $\approx 3$
near the core, meaning that quarks become lighter when they 
approach the inner region of the wall;
for comparison, we checked that for the walls in the phase
with chiral symmetry
broken $\Delta$ decreases of a few percent only moving from the
exterior part of the wall towards the core.

The energy per unit of transverse area, that is the
surface tension $\kappa$,
of the domain wall is defined as
\begin{equation}
\kappa=\int_{-\infty}^{+\infty} dx \left[
\frac{1}{2} \left(\frac{da}{dx}\right)^2
+ V(a/f_a)
\right].
\end{equation}
From the expression above it is easy to see that the 
wall gets most of its energy from the region where
$|da/dx|$ and $V$ are larger, namely for $a/f_\pi=(2k+1)\pi$.
Using $a=f_a\theta$ and Eq.~\eqref{eq:laax4ee}
for $c=0$ we get
\begin{equation}
	\kappa
	=2\sqrt{2}f_a\int_0^{\pi}d\theta\sqrt{V(\theta)},
	\label{eq:po999jjj}
\end{equation}
which stands for both kink and antikink solutions.
For the simple potential~\eqref{eq:largeTpot}
this gives in particular
\begin{equation}
	\kappa = 
	8m_a f_a^2 =\frac{8\chi}{m_a},
	\label{eq:po999jjjAAA}
\end{equation}
where $\chi$ denotes the topological susceptibility.

\begin{figure}[t!]
	\begin{center}
		\includegraphics[scale=0.21]{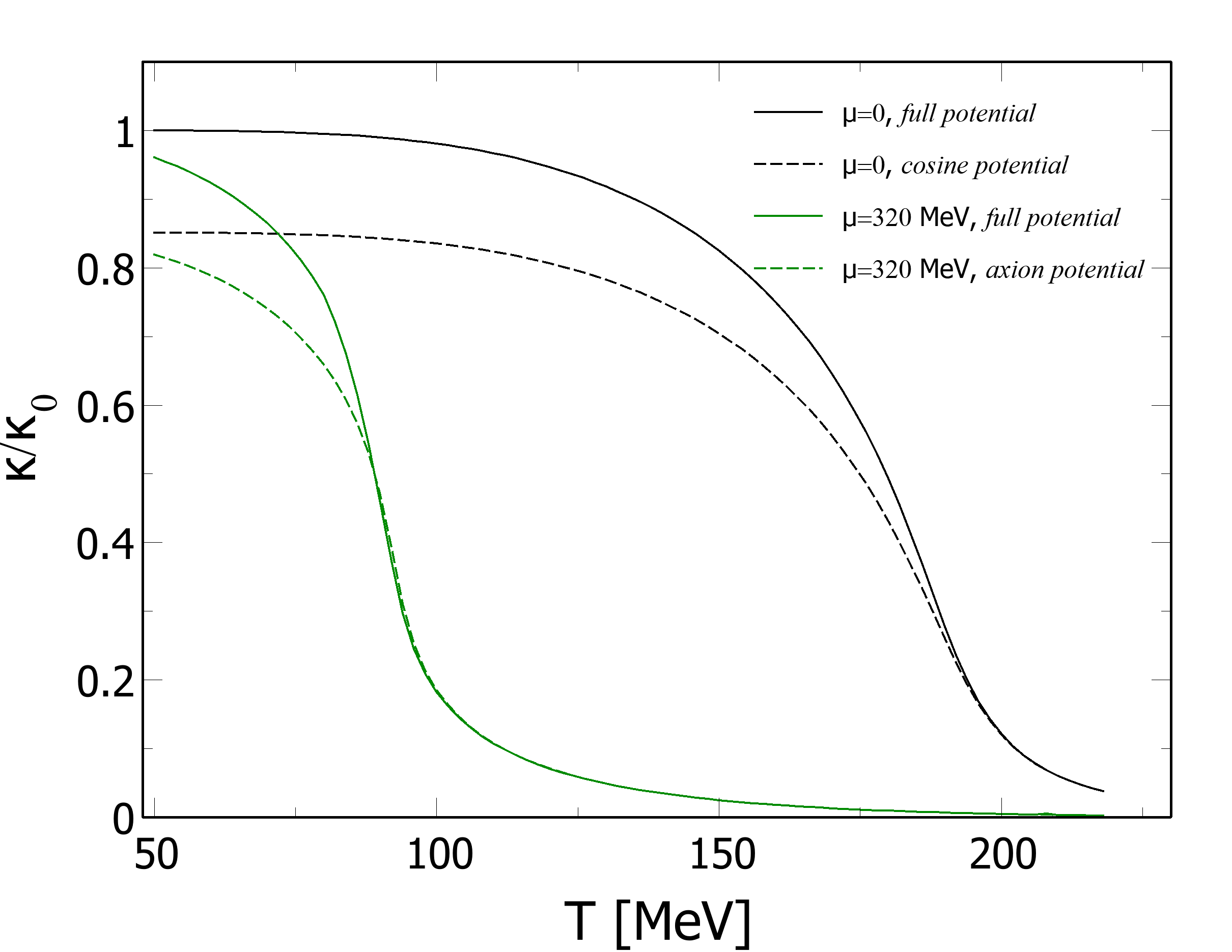}
	\end{center}
	\caption{\label{Fig:kappat}
		$\kappa$ versus temperature at $\mu=0$ (black lines)
and  $\mu=320$ MeV (green lines). Solid lines correspond to the
results obtained with the full potential~\eqref{eq:VdefPPP}, while the dashed lines
are the results obtained by virtue of the simplified
cosine potential~\eqref{eq:largeTpot}. $\kappa_0=1.9\times 10^{16}$ MeV$^3$
is the surface tension computed at $T=\mu=0$.
}
\end{figure}

In Fig.~\ref{Fig:kappat} we plot $\kappa$ versus temperature
for $\mu=0$ (black lines) and
$\mu=320$ MeV (green lines), computed along the neutrality line;
solid lines correspond to the results obtained using the
full axion potential~\eqref{eq:VdefPPP}, while the dashed lines
are the results obtained by virtue of the simplified
cosine potential~\eqref{eq:largeTpot}. 
$\kappa$ is measured in units of the surface tension at
$T=\mu=0$, which is $\kappa_0=1.9\times 10^{16}$ MeV$^3$.
We note that 
the QCD phase transition drastically affects the surface 
tension of the wall;
particularly, in correspondence of chiral restoration $\kappa$ drops
of about one order of magnitude for both values of $\mu$ shown.

We close this section with a comment on the possible abundance of
axion walls in the cores of compact stars.
From Eq.~\eqref{eq:po999jjjAAA} we note that
$\kappa$ can be a monstrous number in the vacuum;
in the presence of dense quark matter, our calculations show that
$\kappa$ can decrease of a few orders of magnitude at most, which
still gives a gigantic surface tension.
Therefore, one might conclude that
forming the walls in quark matter is almost as prohibitive as
forming the walls in the vacuum.
However, this argument does not take into account of the
background energy carried by quark matter itself:
our conclusion is that in the thermodynamic limit,
adding one wall to the bulk quark matter costs zero energy,
therefore axion walls can form easily in presence of
dense quark matter.
To see this effect, for simplicity let us limit ourselves to
the zero temperature case, which is a good approximation
for the core of a compact star. Then, 
taking into account the
domain wall, the energy density at $T=0$ is
\begin{equation}
\mathcal{E}=
\frac{1}{2} \left(\frac{da}{dx}\right)^2
+\Omega(\mu,a(x));
\end{equation}
here $a(x)$ denotes the wall profile,
so $\Omega(\mu,a(x))$ contains both
the contribution of quark matter and that of
the wall. 
We can add and subtract $\Omega(\mu,0)$ to
the right hand side of the above equation,
then subtract the irrelevant constant $\Omega(0,0)$,
to get
\begin{equation}
	\mathcal{E}=
\mathcal{E}_\mathrm{wall} + \mathcal{E}_\mathrm{quarks},
	\label{eq:opm777}
\end{equation}
where
\begin{equation}
\mathcal{E}_\mathrm{wall} = 
\frac{1}{2} \left(\frac{da}{dx}\right)^2
+V[a(x)],
\end{equation}
with $V$ is defined in Eq.~\eqref{eq:VdefPPP},
and
\begin{equation}
\mathcal{E}_\mathrm{quarks}=\Omega(\mu,0)-\Omega(0,0).
\label{eq:quarksYYY}
\end{equation}
Thus, $\mathcal{E}_\mathrm{wall}$ corresponds to the
energy density stored in the wall $a(x)$ at
a given $\mu$, while $\mathcal{E}_\mathrm{quarks}$
is the free energy of bulk quark matter
at the same $\mu$. In other words,
adding the wall $a(x)$ to the
bulk of quark matter requires
an energy density $\mathcal{E}_\mathrm{wall}$.
In the thermodynamic limit, $L\rightarrow\infty$,
the energy of the wall grows $\sim L^2$;
however, the energy of the background of quark
matter grows $\sim \mu^4 L^3$. Accordingly,
the energy cost of adding one of these solitons
to the bulk of quark matter is zero in this limit.
We conclude that forming walls in bulk quark matter is easier
than forming the walls in the vacuum, hence these walls
might be abundant in the cores of compact stellar objects.

\section{Conclusions and outlook} 

We studied the QCD axion potential in dense quark matter.
In particular, we analyzed the axion mass and self-coupling
at finite temperature and/or baryon density. The interaction
of axions to QCD matter, as well as the strong interaction,
were modeled by a local NJL model. Our main goal was to study
the effect of the chiral phase transition on the properties
of the QCD axion.  Interestingly, axions have been studied in astrophysical environments in the context of supernova explosions and protoneutron stars formation~\cite{Lucente:2020whw,Fischer:2021jfm}. Within this scenario, axions or axion like particles might be formed by means of the so called Primakoff process, which involves resonant production of neutral pseudoscalar mesons from the interaction of high-energy photons with atomic nuclei. The expected signals have been searched for in different data surveys, for instance in the Fermi-LAT data, where relevant the energy range covers 50 MeV to 500 GeV~\cite{Calore:2021hhn}. In addition, compact stars can potentially cool down by axion emissions that complement the standard neutrino and photon cooling~\cite{Leinson:2014ioa,Sedrakian:2015krq,Sedrakian:2018kdm,Buschmann:2021juv}. Keeping in mind potential applications of the results
to the astrophysical compact objects, namely neutron stars or
neutron stars mergers, we implemented bulk quark matter 
which is locally electrically neutral. 
We found that the chiral phase transition considerably affects
the low energy properties of axions. 
In fact, the axion mass drops when chiral symmetry
is restored. Moreover, the axion quartic self-coupling
is enhanced when quark matter is close to the QCD critical
endpoint.

We then computed the axion walls in dense quark matter,
focusing on the surface tension of the solitons:
to our knowledge, this is the first time that such a problem
is considered. We noticed that
the energy to form one of such walls in bulk quark matter has to be
compared with the energy of the background matter: in the
thermodynamic limit adding one wall to the bulk costs zero energy.
As a consequence, adding walls to dense quark matter is not disfavored
by energy arguments, and our conclusion is that
it is likely that in bulk quark matter
many axion walls form. 
The calculation of the full axion potential,
in addition with the domain wall tension,
shows that increasing $T$ and/or $\mu$
the potential well of the QCD axion becomes
lower, thus making the transitions between 
the $\theta-$vacua easier.

Differently from
calculations based on $\chi$PT, our work
has at least
two advantages.
Firstly, it gives a result for the full axion potential,
rather than
an expansion around $a=0$.
Moreover, it allows us to take into account the
effect of the chiral phase transition on
the axion potential, and these might have
some impact, see for example the enhancement of 
$\lambda_a$ around the transition, or the 
lowering of $m_a$ in the chiral restored phase.
These effects can not be obtained within
$\chi$PT because in the latter the phase transition
at finite temperature and/or chemical potential is missing.

As previously stated,
this work paves the way to more
complete model calculations as well as to
interesting astrophysical applications. 
The results in Fig. \ref{Fig:qrn2} show that $\lambda_a$
is enhanced in proximity of the QCD critical endpoint:
this implies that when quark matter is close to criticality,
axions self-interaction is enhanced and this might favor
the formation of self-bound axion 
droplets. While in our work this is purely speculative,
this particular problem can be studied in detail
and we aim at addressing it in the near future.
Furthermore, 
the use of nonlocal covariant NJL models is very welcome,
since these allow for a better comparison with
lattice QCD data: studies of the coupling of axions to
dense quark matter within nonlocal models are missing,
so it is of a certain interest to extend the work we presented
here to such models,
possibly including a vector interaction. 
Moreover, it will be interesting to
couple quarks to magnetic, and more generally to electromagnetic,
fields, and study how this coupling affects the low energy 
properties of the axions.
Even more, it is of a great interest to study the modifications
of axions to the QCD equation of state at finite $T$ and $\mu$,
having in mind applications to the structure of compact stars
and neutron stars mergers. 
If axion walls can form inside compact stars, then they
might affect the transport properties of nuclear and quark
matter inside the stars themselves, because of the possible
scatterings of nucleons and/or quarks on the walls.  
We leave these interesting problems to near future works.

\begin{acknowledgements}
M. R. acknowledges John Petrucci for inspiration,
L. Campanelli and
Z. Y. Lu for numerous discussions and S. S. Wan for
his support on the initial part of this project. D.E.A.C. acknowledges support from the NCN OPUS Project No. 2018/29/B/ST2/02576. A.G. G. would like to acknowledge the support received from CONICET (Argentina) under Grants No. PIP 22-24 11220210100150CO and from ANPCyT (Argentina) under Grant No. PICT20-01847.
\end{acknowledgements}

\end{document}